\documentclass{elsart}
\usepackage{graphicx}
\usepackage{amssymb}
\newcommand{\sect}[1]{\setcounter{equation}{0}\section{#1}}

\newcommand{\bfm}[1]{\mbox{\boldmath${#1}$}}
\begin{document}
\begin{frontmatter}
\title{Deformed logarithms and entropies\thanksref{miur}}
\author[INFM]{G. Kaniadakis}
\author[INFN]{M. Lissia}
\author[INFN,INFM]{A.M. Scarfone}
%
\address[INFM]{Istituto Nazionale di Fisica della Materia and
Dipartimento di Fisica,\\ Politecnico di Torino, Corso Duca delgi
Abruzzi 24, I-10129 Torino, Italy}
\address[INFN]{Istituto Nazionale di Fisica Nucleare and Dipartimento di
Fisica,\\ Universit\`a di Cagliari, I-09042 Monserrato, Italy}
\thanks[miur]{This
work was partially supported by MIUR (Ministero dell'Istruzione,
del\-l'Uni\-ver\-si\-t\`a e della Ricerca) under MIUR-PRIN-2003 project
``Theoretical Physics of the Nucleus and the Many-Body Systems''.
}
\date{\today}
\begin {abstract}
By solving a differential-functional equation inposed by the
MaxEnt principle we obtain a class of two-parameter deformed
logarithms and construct the corresponding two-parameter
generalized trace-form entropies. Generalized distributions follow
from these generalized entropies in the same fashion as the
Gaussian distribution follows from the Shannon entropy, which is a
special limiting case of the family. We determine the region of
parameters where the deformed logarithm conserves the most
important properties of the logarithm, and show that important
existing generalizations of the entropy are included as special
cases in this two-parameter class.
\end {abstract}
\end{frontmatter}
\maketitle

\sect{Introduction}
Many experiments in various fields of nuclear and condensed matter
physics suggest the inadequacy of the Boltzmann-Gibbs statistics
and the need for the introduction of new
statistics~\cite{Kaniadakis0,Abe0,Walton,Shlesinger,Landsberg,Frank,Lenzi,Tsallis3,Abe3,Kaniadakis1,Kaniadakis2}.
In particular a large class of phenomena show power-law
distributions with asymptotic long tails. Typically, these
phenomena arise in presence of long-range forces, long-range
memory, or when dynamics evolves in a non-euclidean multi-fractal
space-time. In the last decade many authors pursued new
statistical mechanics theories that mimic the structure of the
Boltzmann-Gibbs theory, while capturing the emerging experimental
anomalous behaviors.

An interesting approach generalizes the Shannon entropy using
suitable modifications of the logarithm. This idea has generated
interesting new entropic forms. Important examples are the
Tsallis-entropy \cite{Tsallis3}, the Abe-entropy \cite{Abe3} and
the $\kappa$-entropy \cite{Kaniadakis1,Kaniadakis2}. Remarkably,
all these entropies belong to a two-parameter class, see Eq.
(\ref{stentropy}), introduced a quarter of century ago by Mittal
\cite{Mittal} and Sharma-Taneja \cite{Taneja1,Taneja2}, and
recently studied by Borges and Roditi \cite{Borges1}.

Let us consider the following class of trace-form entropies (in
this work $k_{\rm B}=1$)
\begin{equation}
S(p)=-\sum_{i=1}^N p_{_i}\,\Lambda(p_{_i}) \ ,\label{defentropy}
\end{equation}
where $p\equiv\{p_{_i}\}_{_{i=1,\cdots,\,N}}$ is a discrete
probability distribution, and $\Lambda(x)$ is an analytical
function that generalizes the logarithm. The main properties of a
very large class of generalized logarithms have been considered in
Refs.~\cite{Naudts1,Naudts2,Naudts3}. The canonical distribution
$p$ is obtained by maximizing the entropy in Eq.
(\ref{defentropy}) for fixed normalization and energy,
\begin{equation}
\sum_{i=1}^N\,p_{_i}=1 \
,\hspace{30mm}\sum_{i=1}^N\,E_{_i}\,p_{_i}=U \ , \label{con}
\end{equation}
obtaining the equation
\begin{equation}
\frac{d}{d\,p_{_j}}\left[p_{_j}\,\Lambda(p_{_j})\right]= -\beta
(E_{_j}-\mu) \ ,\label{eq}
\end{equation}
where $\beta$ and $-\beta\mu$ are the Lagrange multipliers
associated to the constraints (\ref{con}). After introducing the
inverse function of the generalized logarithm, {\it i.e.}, the
generalized exponential ${\mathcal E}(x)=\Lambda^{-1}(x)$, we
require that
\begin{equation}
p_{_j}=\alpha\,{\mathcal
E}\left(-\frac{\beta}{\lambda}\,(E_{_j}-\mu)\right) \
,\label{distribution}
\end{equation}
where $\alpha$ and $\lambda$ are two arbitrary, real and positive
constants. Then Eq. (\ref{eq}) becomes the differential-functional
equation \cite{Kaniadakis2}:
\begin{equation}
\frac{d}{d\,p_{_j}}\left[p_{_j}\,\Lambda(p_{_j})\right]
=\lambda\,\Lambda\left(\frac{p_{_j}}{\alpha}\right) \
.\label{condint}
\end{equation}
The choice $\lambda=1$ and $\alpha=e^{-1}$ yields
an important special case of this class of equations:
the solution of  Eq. (\ref{condint}) with boundary conditions
$\Lambda(1)=0$ and $d\,\Lambda(x)/dx\big|_{x=1}=1$ is
$\Lambda(p_{_j})=\ln p_{_j}$ and entropy  (\ref{defentropy})
reduces to the Shannon-entropy.

Note that, since in general $\Lambda(x)\neq
-\Lambda\left(1/x\right)$, we could use the  different definition
of entropy
\begin{equation}
{\mathcal
S}(p)=\sum_{i=1}^Np_{_i}\,\Lambda\left(\frac{1}{p_{_i}}\right) \
;\label{defentropy1}
\end{equation}
however, we would obtain the same family of deformed logarithms: a
simple mapping of the parameters brings results obtained from Eq.
(\ref{defentropy}) into those  obtained from Eq.
(\ref{defentropy1}).\\ In the present contribution, by solving Eq.
(\ref{condint}), we obtain a class of two-parameter generalized
logarithms which leads to the entropy considered in Refs.
\cite{Mittal,Taneja1,Taneja2,Borges1}.

\sect{The deformed logarithm} Introducing the function $
\Lambda(p_{_j}) \equiv (1 /  p_{_j})\,f(\lambda\,\alpha\,\ln
p_{_j}) $ and performing the change of variable $
p_{_j}=\exp\left(t/(\lambda\,\alpha)\right)$, Eq. (\ref{condint})
becomes an homogeneous differential-difference equation of the
first order, belonging to the class of the delay equations
\cite{Bellman}
\begin{equation}
\frac{d\,f(t)}{d\,t}-f(t-t_{_0})=0 \ ,\label{delay}
\end{equation}
where $ t_{0}=\lambda\,\alpha\,\ln\alpha$.
Its general solution is
\begin{equation}
f(t)=\sum_{i=1}^n\sum_{j=0}^{m_i-1}a_{_{ij}}(s_{_1},\,\cdots,\,s_{_n})\,t^j\,
e^{s_{_i}\,t}\ ,\label{sol}
\end{equation}
with $n$ the number of different solutions $s_{_i}$ of the
characteristic equation
\begin{equation}
s_{_i}-e^{-t_{_0}\,s_{_i}}=0 \ ,\label{char}
\end{equation}
and $m_{_i}$ their multiplicity. In general the integration
constants $a_{_{ij}}$ depend on the parameters $s_{i}$. Eq.
(\ref{char}) admits $n=0,\,1$ or $2$ real solutions, depending on
the value of $ t_{0}$: (a) for $t_{_0}\geq0$, $n=1$ and $m=1$;
(b) for $-1/e<t_{_0}<0$, $n=2$ and $m_{_i}=1$; (c) for
$t_{_0}=-1/e$, $n=1$ and $m=2$ (two coincident solutions);
finally, for $t_{_0}<-1/e$ there exists no real solution. The
case (a) gives the trivial solution
\begin{equation}
\Lambda(p_{_i})=a\,p_{_i}^b \ ,
\end{equation}
with $b=\lambda\,\alpha\,s-1$. Clearly, this solution can not be
used to define a generalized logarithm.

The general solution for case (b) is
\begin{equation}
\Lambda(p_{_i})=A_{_1}(\kappa_{_1},\,\kappa_{_2})\,
p_{_i}^{\kappa_{_1}}+A_{_2}(\kappa_{_1},\,\kappa_{_2})\,
p_{_i}^{\kappa_{_2}}\ ,\label{sol1}
\end{equation}
where $\kappa_{_i}=\lambda\,\alpha\,s_{_i}-1$ and
$A_{_i}(\kappa_{_1},\,\kappa_{_2})$ are integration constants. The
boundary conditions
\begin{equation} \Lambda(1)=0 \
,\hspace{10mm}{\rm and}\hspace{10mm}
\frac{d\,\Lambda(x)}{d\,p_{_i}}\Bigg|_{p_{_i}=1}=1 \
,\hspace{10mm}\forall\, \kappa_{_1},\,\kappa_{_2} \ ,\label{norm}
\end{equation}
imply, respectively, $A_{_1}=-A_{_2}$ and $\kappa_{_1}\,A_{_1}+
\kappa_{_2}\,A_{_2}=(\kappa_{_1}-\kappa_{_2})\,A_{_1}=1$; then
Eq. (\ref{sol1}) becomes
\begin{equation}
\Lambda(p_{_i})=\frac{p_{_i}^{\kappa_{_1}}-p_{_i}^{\kappa_{_2}}}{\kappa_{_1}
-\kappa_{_2}} \ .\label{log1}
\end{equation}
In the following we introduce a different parametrization
$\kappa=(\kappa_{_1}-\kappa_{_2})/2$ and
$r=(\kappa_{_1}+\kappa_{_2})/2$ and write Eq. (\ref{norm}) as:
\begin{equation}
\ln_{_{\{\kappa,\,r\}}}(p_{_i})=p_{_i}^r\,
{\displaystyle\frac{p_{_i}^\kappa-p_{_i}^{-\kappa}}{2\,\kappa}} \
,\label{log}
\end{equation}
where the notation
$\ln_{_{\{\kappa,r\}}}(p_{_i})\equiv\Lambda(p_{_i})$ has been
used. The boundary condition $\Lambda(0^+)<0$ implies
$r\leq|\kappa|$. From the characteristic equation (\ref{char}) we
obtain the system
\begin{equation}
\left\{
\begin{array}{l}
1+r+\kappa=\lambda\,\alpha^{-r-\kappa} \
,\\
1+r-\kappa=\lambda\,\alpha^{-r+\kappa} \ ,\label{condition}
\end{array}
\right.
\end{equation}
which can be solved for the two constants $\alpha$ and $\lambda$
\begin{equation}
\alpha=\left(\frac{1+r-\kappa}{1+r+\kappa}\right)^{1/2\,\kappa} \
,\hspace{20mm}
\lambda=\frac{\left(1+r-\kappa\right)}{\mbox{\raisebox{-1mm}
{$\left(1+r+\kappa\right)$}}}
^{\scriptscriptstyle{(r+\kappa)/2\,\kappa}}_
{\mbox{\raisebox{3.5mm}{$\scriptscriptstyle{(r-\kappa)/2\,\kappa}$}}}
 \ .\label{l}
\end{equation}
In the following, we call the solution (\ref{log}) deformed or
$(\kappa,\,r)$-logarithm. We remark that Eqs. (\ref{log}) and
(\ref{l}) are invariant under $\kappa \to -\kappa$ and that Eq.
(\ref{log}) reduces to the standard logarithm in the
$(\kappa,\,r)\rightarrow(0,\,0)$ limit.

Finally, we discuss case (c). This case corresponds to the limit
of case (b) $\kappa_{_1}\to \kappa_{_2}$ and then, from Eq.
(\ref{log}) we obtain
\begin{equation}
\ln_{_{\{0,\,r\}}}(p_{_i})=p_{_i}^r\,\ln p_{_i} \ ,\label{part}
\end{equation}
which does not satisfy the condition $\ln_{_{\{0,\,r\}}}(0^+)<0$.

\sect{Properties of the $\bfm{(\kappa,\,r)}$-logarithm}

Naudts \cite{Naudts1} gives a list of properties
that a deformed logarithm must satisfy for the corresponding
entropy to be physical. In this Section we determine the region of
the parameter space $(\kappa, r)$ where the logarithm (\ref{log})
verifies the following properties:
\begin{eqnarray}
&&\ln_{_{\{{\scriptstyle \kappa,\,r}\}}}(x) \in C^{\infty}(I\!\!
R^+) \ ,
\label{pd1}\\
&&\frac{d}{d\,x}\, \ln_{_{\{{\scriptstyle \kappa,\,r}\}}}(x)>0 \
,\hspace{32.5mm}-|\kappa|\leq r\leq|\kappa| \ ,
\label{pd2}\\
&&\frac{d^2}{d\,x^2}\, \ln_{_{\{{\scriptstyle
\kappa,\,r}\}}}(x)<0 \ ,\hspace{30mm}\, -|\kappa|\leq
r\leq\frac{1}{2}-\Big|\frac{1}{2}-|\kappa|\Big| \ ,
\label{pd3}\\
&&\ln_{_{\{{\scriptstyle \kappa,\,r}\}}}(1)=0 \ , \label{pd4}\\
&&\int\limits_0\limits^1\ln_{_{\{\kappa,\,r\}}}(x)\,dx=-\frac{1}{(1+r)^2
-\kappa^2} \ ,\hspace{8mm}1+r>|\kappa| \ ,\label{pd5}\\
&&\int\limits_0\limits^1\ln_{_{\{\kappa,\,r\}}}\left({1\over
x}\right)\,dx=\frac{1}{(1-r)^2 -\kappa^2} \
,\hspace{7.5mm}1-r>|\kappa| \ .\label{pd51}
\end{eqnarray}
The $(\kappa,\,r)$-logarithm is an analytical function for all
$x\geq0$ and for all $\kappa,\,r\in I\!\!R$, (\ref{pd1}); is a
strictly increasing function for $-|\kappa|\leq r\leq|\kappa|$,
(\ref{pd2}); is concave for $-|\kappa|\leq r\leq 1/2 -
|1/2-|\kappa|\,| $, (\ref{pd3}); verifies the boundary conditions
(\ref{norm}); has at most integrable divergences at $x=0$ and
$x=+\infty$, (\ref{pd5}) and (\ref{pd51}). All these conditions
(\ref{pd2})--(\ref{pd51}) select the region of the parameter space
\begin{eqnarray}
I\!\!R^2\supset{\mathcal R}=\left\{
\begin{array}{l}
\hspace{4.5mm}-|\kappa|\leq r\leq|\kappa|\hspace{27mm} {\rm if} \
0\leq|\kappa|<{1\over2} \ ,\\
|\kappa|-1<r<1-|\kappa|\hspace{20mm} {\rm if} \
{1\over2}\leq|\kappa|<1 \ .
\end{array}
\right. \quad , \label{quadri}
\end{eqnarray}
which is shown in Fig.~\ref{fig:parameterRegion}. Let us remark
that this region includes value of the parameters for which the
deformed logarithm is finite for $x\rightarrow0^+$ or
$x\rightarrow+\infty$, that $\kappa\rightarrow0$ implies
$r\rightarrow0$, and that the restriction $|\kappa|<1$ was
already obtained in Ref. \cite{Kaniadakis2} for the case $r=0$.

The asymptotic behaviors of $\ln_{_{\{\kappa,r\}}}(x)$, for $x\to
0^+$ and $x\to+\infty$ are
\begin{equation}
\ln_{_{\{\kappa,r\}}}(x){\atop\stackrel{\textstyle\sim}{\scriptstyle
x\rightarrow
{0^+}}}-\frac{1}{2\,|\kappa|}\,\frac{1}{x^{|\kappa|-r}} \quad\quad
\ln_{_{\{\kappa,r\}}}(x){\atop\stackrel{\textstyle\sim}{\scriptstyle
x\rightarrow {+\infty}}}\frac{1}{2\,|\kappa|}\,x^{|\kappa|+r} \
; \label{pd8}
\end{equation}
these divergences, which are integrable inside the
region (\ref{quadri}), become finite values
for specific choices on its boundary.

Finally, note that the property of the  standard logarithm
$\ln (1/x)=-\ln x$ is satisfied by the $(\kappa,\,r)$-logarithm
only for $r=0$ \cite{Kaniadakis2}, otherwise
\begin{equation}
\ln_{_{\{\kappa,\,r\}}}\left(\frac{1}{x}\right)=-\ln_{_{\{\kappa,\,-r\}}}(x)
\ ,\label{pd6}
\end{equation}
showing that the entropies (\ref{defentropy1}) are the same as the
ones defined by Eq. (\ref{defentropy}) with the re-parametrization
$r\rightarrow-r$.

Being the deformed logarithm (\ref{log}) a strictly
monotonic function  for $\kappa,\,r\in{\mathcal R}$
(\ref{pd2}), its inverse function exists and we call
it deformed exponential
$\exp_{_{\{{\scriptstyle \kappa,\,r}\}}}(x)$.
Its analytical properties readily follow from the ones
of the deformed logarithm (\ref{pd1})--(\ref{pd51}):
\begin{eqnarray}
&&\exp_{_{\{{\scriptstyle \kappa,\,r}\}}}(x) \in C^{\infty}(I\!\!
R) \ ,\hspace{27mm}\kappa,\,r\in{\mathcal R}-\{r=\pm|\kappa|\} \ ,
\label{ed1}\\
&&\frac{d}{d\,x}\, \exp_{_{\{{\scriptstyle \kappa,\,r}\}}}(x)>0 \
,\hspace{32mm}-|\kappa|\leq r\leq|\kappa| \ ,
\label{ed2}\\
&&\frac{d^2}{d\,x^2}\, \exp_{_{\{{\scriptstyle
\kappa,\,r}\}}}(x)>0 \ ,\hspace{29.5mm}\, -|\kappa|\leq
r\leq{1\over2}-\Big|{1\over2}-|\kappa|\Big| \ ,\label{ed3}\\
&&\exp_{_{\{{\scriptstyle
\kappa,\,r}\}}}(0)=1 \ ,
\label{ed4}\\
&&\int\limits_{-\infty}\limits^0\exp_{_{\{\kappa,\,r\}}}(x)\,dx=\frac{1}
{(1+r)^2-\kappa^2} \ ,\hspace{7mm}1+r>|\kappa| \ ,\label{ed5}\\
&&\int\limits_{-\infty}\limits^0{dx\over\exp_{_{\{\kappa,\,r\}}}(-x)}=\frac{1}
{(1-r)^2-\kappa^2} \ ,\hspace{8mm}1-r>|\kappa| \ .\label{ed51}
\end{eqnarray}
Note that Eq. (\ref{ed1}) states that for choices $r=\pm|\kappa|$
the deformed exponential is defined on a reduced domain  $x\in
{\rm Image}[\log_{_{\{{\scriptstyle
\kappa,\,\pm|\kappa|}\}}}(x)]$ and not  for all $x\in I\!\!R$,
since the  deformed logarithm goes to a finite value, $\mp
1/|2\kappa|$, when $x\to 0^+$ or $x\to +\infty$. Property
(\ref{pd6}) implies
\begin{equation}
\exp_{_{\{\kappa,\,r\}}}(x)\,\exp_{_{\{\kappa,\,-r\}}}(-x)=1 \
,\label{ed76}
\end{equation}
which coincides with the one satisfied by the standard exponential,
$\exp(x)\,\exp(-x)=1$, only for $r=0$. \cite{Kaniadakis2}.

Finally, the asymptotic power-law behaviors of
$\exp_{_{\{\kappa,r\}}}(x)$ are
\begin{equation}
\exp_{_{\{\kappa,r\}}}(x){\atop\stackrel{\textstyle\sim}{\scriptstyle
x\rightarrow {\pm\infty}}}|2\,\kappa\,x|^{1/(r\pm|\kappa|)} \
.\label{expasy}
\end{equation}

\sect{Special examples of deformed logarithms} Different authors
have introduced one-parameter families of logarithms in the
context of generalized statistical mechanics. In the following we
show how some important instances belong the two-parameter class
under exam for suitable choices of the parameters.

\subsection{Tsallis' logarithm}

If we impose $r=\pm|\kappa|$ (dashed lines in
Fig.~\ref{fig:parameterRegion}) and introduce the parameter
$q=1\mp2\,|\kappa|$, Eq. (\ref{log}) becomes Tsallis' logarithm
\cite{Tsallis3}
\begin{equation}
\ln_q\,(x)=\frac{x^{1-q}-1}{1-q} \ ,\label{tlog}
\end{equation}
which is an  analytical,  increasing and concave function for all
$x\geq0$; property (\ref{pd5}) becomes
$\int_0^1\ln_q\,(x)\,dx=-1/(2-q)$. The relation (\ref{pd6})
becomes
\begin{equation}
\ln_q\,\left(\frac{1}{x}\right)=-\ln_{2-q}\,(x) \
.\label{condition1}
\end{equation}

The asymptotic behaviors are
\begin{equation}
\ln_q(x){\atop\stackrel{\textstyle\sim}{\scriptstyle x\rightarrow
{0^+}}}-\frac{1}{1-q} \quad\quad \mathrm{and} \quad\quad
\lim_{x\to +\infty} \ln_q(x) = \frac{x^{1-q}}{1-q}
\end{equation}
for $q\in(0,\,1)$; behaviors for $q\in(1,\,2)$ follow from Eq.
(\ref{condition1}).
%

We remark that the relation between Tsallis' logarithm
and entropy is obtained from Eq. (\ref{defentropy})
with the substitution $q\rightarrow2-q$.

The $q$-exponential is
\begin{equation}
\exp_q\,(x)=[1+(1-q)\,x]^{1/(1-q)} \ ,\label{qexp}
\end{equation}
which is an analytic, monotonic and convex function in
$x\in[-1/(1-q),\,+\infty)$ for $q\in(0,\,1)$ and in
$x\in(-\infty,\,1/(q-1)]$ for $q\in(1,\,2)$. When $q\in(0,\,1)$,
the exponential (\ref{qexp}) diverges as
\begin{equation}
\exp_q(x){\atop\stackrel{\textstyle\sim}{\scriptstyle
x\rightarrow +\infty}}[(1-q)\,x]^{1/(1-q)} \ ,
\end{equation}
while it approaches to zero for $x\rightarrow-1/(1-q)$. As for
the logarithm, the behavior for $q\in(1,\,2)$ can be obtained from
the symmetry $(x,q)\to (-x,2-q)$.

\subsection{Abe logarithm}
The general logarithm  (\ref{log}) with the constraint
$r=\sqrt{1+\kappa^2}-1$ (dash-dotted line in
Fig.~\ref{fig:parameterRegion}) becomes the logarithm associated
to the entropy introduced by Abe \cite{Abe3}
\begin{equation}
\ln_{q_{_{\rm A}}}\,(x)=\frac{x^{(q_{_{\rm
A}}^{-1})-1}-x^{q_{_{\rm A}}-1}}{{ q_{_{\rm A}}}^{-1}-q_{_{\rm
A}}} \ ,\label{abelog}
\end{equation}
where $q_{_{\rm A}}=\sqrt{1+\kappa^2}+\kappa$; the standard
logarithm is recovered for $q_{_{\rm A}}\rightarrow1$. The fact
that the Abe-logarithm is a special case of a two-parameter
generalized logarithm has already been noticed \cite{Borges1}.\\
The logarithm (\ref{abelog}) is analytic, monotonic and concave
for $q_{_{\rm A}}\in(1/2,\,2)$ and $x\geq0$; it is invariant under
$q_{_{\rm A}}\to1/q_{_{\rm A}}$. Eq. (\ref{pd5}) becomes
$\int_0^1\ln_{q_{_{\rm A}}}\,(x)\,dx=-1$, while its asymptotic
behaviors for $q_{_{\rm A}}<1$ are
\begin{equation}
\ln_{q_{_{\rm
A}}}\,(x)\,\,{\atop\stackrel{\textstyle\sim}{\scriptstyle
x\rightarrow {0^+}}}-\frac{x^{q_{_{\rm A}}-1}}{q_{_{\rm
A}}^{-1}-q_{_{\rm A}}} \ ,\hspace{10mm} \ln_{q_{_{\rm
A}}}\,(x)\,\,{\atop\stackrel{\textstyle\sim}{\scriptstyle
x\rightarrow {+\infty}}}\,\,\frac{x^{(q_{_{\rm
A}}^{-1})-1}}{{q_{_{\rm A}}^{-1}-q_{_{\rm A}}}} \ .
\end{equation}
The corresponding inverse function, the Abe exponential, has all
the general properties (\ref{ed1})--(\ref{ed51}) plus the
specific ones corresponding to the symmetries of the logarithm. It
is not possible to express this inverse in terms of known
functions.

\subsection{$\kappa$-logarithm}

\begin{figure}[h]
\hspace{-20mm}
\includegraphics[width=1.2\textwidth]{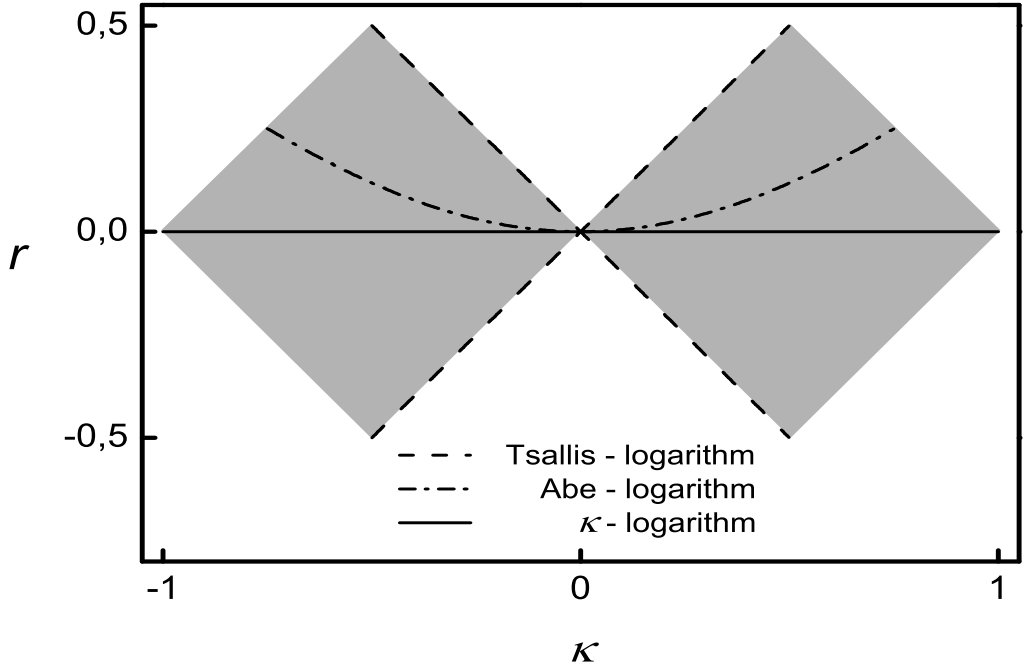} \caption{Parameter space
$(\kappa, r)$ for the logarithm (\ref{log}). The shaded region
represents the constraints of Eq. (\ref{quadri}) on the
parameters. The three lines, dashed, dash-dotted and solid,
correspond to the Tsallis-, Abe- and $\kappa$-logarithm,
respectively, defined by Eqs. (\ref{tlog}), (\ref{abelog}), and
(\ref{klog}). \label{fig:parameterRegion} }
\vspace{10mm}
\end{figure}

Given the form (\ref{log}) of the logarithms we are studying, it
is of special interest the symmetric choice  $r=0$ (solid line in
Fig.~\ref{fig:parameterRegion}) which yields the
$\kappa$-logarithm introduced in Ref. \cite{Kaniadakis1}
\begin{equation}
\ln_{_{\{\kappa\}}}(x)=\frac{x^\kappa-x^{-\kappa}}{2\,\kappa} \
\quad\quad\quad -1<\kappa < 1 \; .\label{klog}
\end{equation}
In fact this $\kappa$-logarithm  is the only member of the
family satisfying the relation
\begin{equation}
\ln_{_{\{\kappa\}}}\left(\frac{1}{x}\right)=-\ln_{_{\{\kappa\}}}(x)
\ ,\label{kd}
\end{equation}
making equivalent the choices $-\log_{_{\{\kappa\}}}(x)$ or
$\log_{_{\{\kappa\}}}(1/x)$ for constructing the entropy. The main
properties of $\ln_{{\{\kappa\}}}(x)$ are those in Eqs.
(\ref{pd1})--(\ref{pd51}); in particular
$\int_0^1\ln_{_{\{\kappa\}}}(x)\,dx=-1/(1-\kappa^2)$, and the
asymptotic behaviors are
\begin{equation}
\ln_{_{\{\kappa\}}}(x)\,\,{\atop\stackrel{\textstyle\sim}{\scriptstyle
x\rightarrow {0^+}}}-{1\over2\,|\kappa|}\,{1\over x^{|\kappa|}} \
,\hspace{10mm}
\ln_{_{\{\kappa\}}}(x)\,{\atop\stackrel{\textstyle\sim}{\scriptstyle
x\rightarrow {+\infty}}}\,{1\over2\,|\kappa|}\,x^{|\kappa|} \ .
\end{equation}
The inverse function of $\ln_{_{\{\kappa\}}}(x)$ is the
$\kappa$-exponential:
\begin{equation}
\exp_{_{\{\kappa\}}}(x)=\left(\sqrt{1+\kappa^2\,x^2}+\kappa\,x
\right)^{1/\kappa}\ ,\label{kexp}
\end{equation}
which is an analytical, monotonic and convex function for all
$x\in{I\!\!R}$ that reproduces the standard exponential for
$\kappa\to 0$ and that has the asymptotic power-law behaviors
\begin{equation}
\exp_{_{\{\kappa\}}}(x)\,{\atop\stackrel{\textstyle\sim}
{\scriptstyle x\rightarrow {\pm\infty}}}\,|2\,\kappa
\,x|^{\pm|\kappa|} \ .
\end{equation}
In addition, it verifies the relation
\begin{equation}
\exp_{_{\{\kappa\}}}(x)\,\exp_{_{\{\kappa\}}}(-x)=1 \ :
\end{equation}
no ambiguity exists between  $\exp_{_{\{\kappa\}}}(-x)$ or
$1/\exp_{_{\{\kappa\}}}(x)$ as statistical weight.

\sect{Conclusions} We have studied the bi-parametric family of
deformed logarithms (\ref{log}). These logarithms are obtained by
solving the differential-functional equation (\ref{condint})
arising from the canonical MaxEnt principle with the resulting
entropy
\begin{equation}
\label{stentropy} S_{\kappa,\,r}(p)=-\sum_{i=1}^N p_{_i}^{1+r}
\,\frac{p_{_i}^\kappa-p_{_i}^{-\kappa}}{2\,\kappa} \quad .
\end{equation}
We have determined the ranges for the deformed parameters
$\kappa$ and $r$ (see Fig.~\ref{fig:parameterRegion}) for
which the $(\kappa,\,r)$-logarithm verifies those properties
that allow Eq. (\ref{stentropy}) to be used as entropy.

Several important one-parameter generalized entropies
(Tsallis-entropy, Abe-entropy and $\kappa$-entropy) have been
shown to belong to this family. There remains the question of the
relevance of each mathematically sound entropy to specific
physical situations.


\end{document}